\documentclass[aps,prl,reprint]{revtex4-1}

\usepackage{amsmath}
\usepackage{graphicx} 
\usepackage{xcolor}
\usepackage{booktabs}
\usepackage{hyperref}
\hypersetup{colorlinks=false, citebordercolor=green}
\usepackage{soul}

\begin{document}
\title{Polarons and their induced interactions in highly imbalanced triple mixtures}

\author{Kevin Keiler $^1$}
\author{Simeon I. Mistakidis $^1$}
\author{Peter Schmelcher $^{1,2}$}
\affiliation{$^1$Center for Optical Quantum Technologies, Department of Physics, University of Hamburg, Luruper Chaussee 149, 22761 Hamburg, Germany}
\affiliation{$^2$The Hamburg Centre for Ultrafast Imaging, University of Hamburg, Luruper Chaussee 149, 22761 Hamburg, Germany}

\begin{abstract}
	We unravel the polaronic properties of impurities immersed in a correlated trapped one-dimensional (1D) Bose-Bose mixture. This setup allows for the impurities to couple either attractively or repulsively to a specific host, thus offering a highly flexible platform for steering the emergent polaronic properties. Specifically, the impurity residue peak and strength of induced interactions can be controlled by varying the coupling of the impurities to the individual bosonic components. In particular, it is possible to maintain the quasiparticle character for larger interaction strengths as compared to the case of impurities immersed in a single bosonic species. We explicate a hierarchy of the polaron binding energies in terms of the impurity-medium interactions, thereby elucidating the identification of the polaronic resonances in recent experimental radiofrequency schemes. 
	For strong attractive impurity-medium couplings bipolaron formation is captured.
	Our findings pave the way for continuously changing the quasiparticle character, under the impact of trap effects, while exposing the role of correlations in triple mixture settings.
\end{abstract}

\maketitle

\paragraph{\label{sec:introduction}Introduction.-}
Ultracold atoms provide pristine platforms for probing quantum phenomena in multi-component fermionic and bosonic \cite{Inouye,Fukuhara_mixt} settings offering an exquisite tunability \cite{feshbach,Inouye,selim}. Highly particle imbalanced mixtures \cite{massignan_review,Kohstall,Koschorreck,Scazza} have lately received major attention in terms of the quasiparticle context \cite{Landau}, leading to fundamentally new insights concerning Fermi and Bose polarons \cite{Schmidt_rev,massignan_review,Grusdt_rev} and thus serving as a quantum simulator of the corresponding condensed matter setup. The quasiparticle notion extends far beyond cold atom settings in semiconducting \cite{cond1} and superconducting devices \cite{cond2}, while interactions among quasiparticles in liquid Helium mixtures \cite{cond3,cond4} and cuprates \cite{cond5,cond6} are a promising candidate for conventional and high-$T_c$ superconductivity \cite{cond7,cond8,cond9,cond10,cond11,cond12,cond13}. Owing to the recent experimental realization of these impurity systems \cite{catani,fukuhura,jorgensen,hu,yan,skou,Kohstall,Koschorreck,Scazza}, an intense theoretical activity has been triggered for the investigation of their stationary properties \cite{volosniev,Sacha_loc} e.g. unveiling their effective mass \cite{grusdt,Khandekar,Ardila_mass}, excitation spectra \cite{Koschorreck,Schmidt_rev,Cetina_intef,Tajima_spec} and induced-interactions \cite{naidon_ind,lew_ind,simosfermi,keiler1,keiler2,jie_chen,zinner_ind,bipolaron1,bipolaron2,bipolaron3,bipolaron4}.
Only very recently, extensions to impurities interacting with a coherently coupled two-component Bose-Einstein condensate (BEC) \cite{coherent_BEC1,coherent_BEC2} and Bose-Bose mixtures \cite{polarons_mixture} have been considered, while introducing holes into spinor fermionic lattice setups led to the concept of magnetic polarons \cite{magnetic_polaron3,magnetic_polaron1,magnetic_polaron2}.
The description of such systems is expected to necessitate higher-order correlations \cite{higherorder5,higherorder1,higherorder2,higherorder3,higherorder4}, thus invalidating lower-order approaches \cite{mean1,mean2,mean3,mean4,polarons_mixture,fro1,fro2,fro3,fro4,fro5,beyofro1,beyofro2,beyofro3,beyofro4,beyofro5,beyofro6}, as already demonstrated in binary systems.

The generalization to triple mixture settings allows for the impurities to selectively couple to the individual hosts, thus offering an efficient platform for tuning the emergent polaronic properties. This includes the longevity, i.e. the prevention of orthogonality catastrophe, the mobility of the polarons and the control of their induced interactions. Additionally, it enables the design of their magnetic and spin-mixing processes by the use of Raman coupling and the design of intriguing bound states such as dimers and trimers as shown for an impurity in a double Fermi sea \cite{double_fermi_sea}. 
Being experimentally within reach of current state-of-the-art experiments \cite{tm_exp1,tm_exp2}, triple mixtures will exhibit more complex quantum phases as compared to their binary counterpart. In this sense, the possibility for the cumulative bath to be prepared in the well-known corresponding phases of a binary mixture, such as in any combination of Tonks-Girardeau gases, miscible phase as well as in an immiscible phase \cite{binary_phases_march}, will naturally impact the quasiparticle character. This will shed light on the polaron problem from a very different perspective as the impurities are dressed by the excitations of two different hosts. Utilizing an ab-initio approach, as we do here, it is possible to enter unexplored regimes where correlations are dominant, thereby serving in particular as a benchmark for future experimental implementations of triple mixture setups as well as  effective theoretical models.  
%Moreover, bipolarons, whose formation stems from the induced interactions, play also a crucial role in technologically relevant materials {\cite{material1}}, e.g. GaAs, and their applications including electric conductivity of polymers\cite{polymer1,polymer2,polymer3} and organic magneto-resistance \cite{organic_magneto}. As such, control over the formation of polarons and bipolarons, besides improving our understanding of the quasiparticle character on a fundamental level, has the potential for designing material properties \cite{material2,material3} and can be useful for photonic and photochemistry applications \cite{material1,material4}.

For these reasons, in this Letter, we undertake the initial step in this direction and explore the polaronic properties of impurities coupled to a 1D \cite{few,few1} harmonically trapped Bose-Bose mixture spanning a wide range of attractive and repulsive impurity-medium coupling strengths, while including all particle correlations. Commonly polarons are studied in spatially uniform systems, while we account for trap effects which are relevant to typical ultracold atom experiments. We exemplify that for a single impurity, the distribution of the impurity residue in terms of the impurity-medium couplings can be steered by adjusting the different interactions to the respective bath of the mixture. In particular, when coupling to the one host repulsively and attractively to the other one, the residue peak can be broadened, such that the polaronic character is maintained for larger interaction strengths. For strong repulsive or attractive impurity-medium interactions the impurity residue vanishes. The behavior of the dressed impurity can be intuitively interpreted in terms of an effective potential, which provides a good approximation for weak impurity-bath coupling strengths, where interspecies entanglement is suppressed. The location of the attractive and repulsive quasiparticle resonances is captured by monitoring the polaron binding energy.
Upon considering two bosonic impurities, we identify the presence of attractive induced interactions whose strength can be steered by the coupling to the respective host of the Bose-Bose mixture. Induced interactions strongly influence the impurities' spatial distribution allowing, for instance, bipolaron formation and lead to a reduction of the impurity residue.

\paragraph{\label{sec:model}Model.-}
We consider a Bose-Bose mixture consisting of two species A and B with equal masses $m_A=m_B=m$ and $N_A=N_B=10$ particles. We note that our results persist for larger $N_A,N_B$, see \cite{supplemental_material}. $N_C=1,2$ bosonic impurities of mass $m_C$ are immersed in this 1D harmonically confined \cite{boshier,Grimm_trap} mixture of interacting atoms. The trap frequencies are $\omega_A=\omega_B=\omega_C=\omega=1.0$. The many-body (MB) Hamiltonian of the system reads

\begin{equation}
\hat{H}=\sum_{\sigma\in\{A,B,C\}}\hat{H_\sigma}+\hat{H}_{AB}+\hat{H}_{AC}+\hat{H}_{BC}.
\label{eq:hamiltonian}
\end{equation}
Here, $\hat{H}_\sigma=\int dx \hat{\Psi}_{\sigma}^{\dagger}(x) ( -\frac{\hbar^{2}}{2 m_\sigma} \frac{d^{2}}{dx^{2}}+ \frac{1}{2}m_\sigma \omega^{2}_\sigma x^2)\hat{\Psi}_{\sigma}(x)+ g_{\sigma\sigma} \int dx \hat{\Psi}_{\sigma}^{\dagger}(x) \hat{\Psi}_{\sigma}^{\dagger}(x) \hat{\Psi}_{\sigma}(x) \hat{\Psi}_{\sigma}(x)$ describes the Hamiltonian of species $\sigma\in \{A,B,C\}$, with contact intraspecies interaction of $g_{AA}=g_{BB}>0$ and $g_{CC}=0$. $\hat{\Psi}_{\sigma}(x)$ is the $\sigma$-species bosonic field operator. $\hat{H}_{\sigma \sigma^{'}}=g_{\sigma \sigma^{'}}\int dx  \hat{\Psi}_{\sigma }^{\dagger}(x) \hat{\Psi}_{\sigma }(x) \hat{\Psi}_{\sigma^{'}}^{\dagger}(x) \hat{\Psi}_{\sigma^{'}}(x)$ denotes the contact interspecies interaction of strength $g_{\sigma \sigma^{'}}$ \cite{olshanii}. In this sense, $\hat{H_A}+\hat{H_B}+\hat{H}_{AB}$ build the Bose-Bose mixture serving as a cumulative bath for the impurity species, described by $\hat{H}_C$. The impurities couple repulsively or attractively to both A and B hosts via a contact interaction of strength $g_{AC}$ and $g_{BC}$, as captured by $\hat{H}_{AC}$ and $\hat{H}_{BC}$. To directly expose the pure effect of impurity-impurity induced interactions we set $g_{CC}=0$. We focus on the case of equal masses $m=m_C$, which can be experimentally realized to a good approximation by considering a mixture of isotopes, e.g. a $^{87}\rm{Rb}$ BEC where the Bose-Bose mixture refers to two hyperfine states \cite{exp_spec1,exp_spec2} and $^{85}\rm{Rb}$ for the impurities.
The effects of mass-imbalance are discussed in Ref. \cite{supplemental_material}.
Throughout this work, we consider $g_{AA}=g_{BB}=0.2$ and $g_{AB}=0.1$ in units of $\sqrt{\hbar^3\omega/m}$, leading to a miscible mixture of species A and B. Spatial scales are given in harmonic units of $\sqrt{\hbar/m\omega}$ and energies in terms of $\hbar\omega$.
To address the ground state of our three-component system we use the variational Multi-Layer Multi-Configuration Time-Dependent Hartree method for atomic mixtures (ML-MCTDHX) \cite{mlb1,mlb2,mlx}. This non-perturbative approach relies on expanding the MB wavefunction with respect to a variationally optimized time-dependent basis \cite{supplemental_material}.

\begin{figure}[t]
	\includegraphics[scale=0.16]{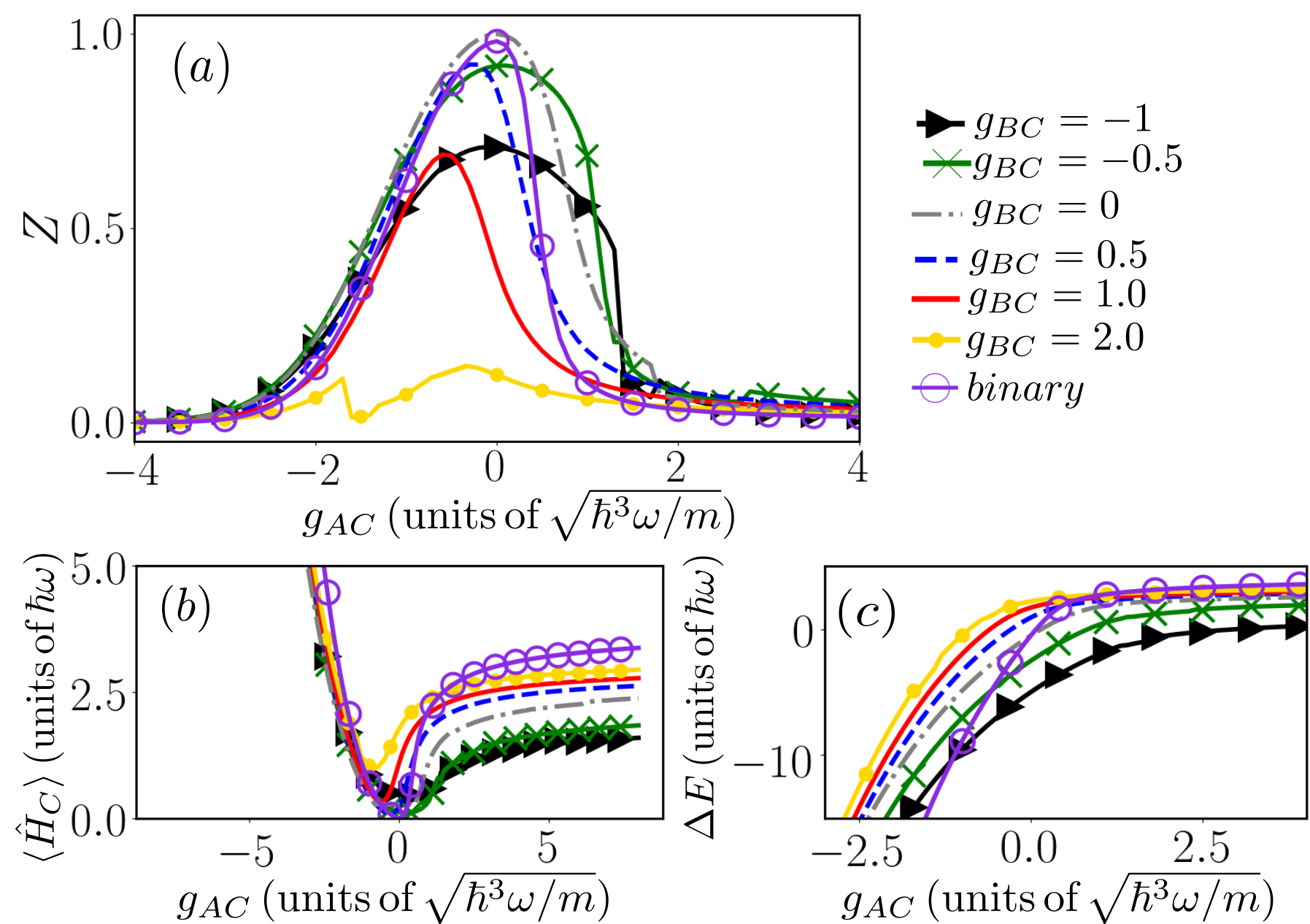}
	\caption{\label{fig:residue_N10} (a) Impurity residue $Z$, polaron (b) energy $\langle\hat{H}_C\rangle$ and (c) binding energy $\Delta E$ for different impurity-bath couplings $g_{AC}$ and $g_{BC}$. The violet circles represent the binary mixture with $N=20$ single species bath atoms and $g=0.2$.}
\end{figure}

\paragraph{\label{sec:results}Results and discussion.-}
As a first step, we vary the individual impurity-medium coupling strengths $g_{AC}$ and $g_{BC}$ of a single impurity to the cumulative bath from attractive to repulsive values and obtain the ground state of the triple mixture. The underlying impurity residue $Z$ \cite{massignan_review} is determined by
\begin{equation}
Z=|\langle\Psi_0|\Psi\rangle|^2,
\label{eq:resdiue}
\end{equation}
where $|\Psi_0\rangle$ is the MB wavefunction for a non-interacting impurity with $g_{AC}=g_{BC}=0$, while $|\Psi\rangle$ denotes the interacting case. Additionally, we determine the impurity residue for a binary mixture, which we define here as an impurity immersed into a bath of $N=20$ single species bosons interacting repulsively with a strength of $g=0.2$. Figure \ref{fig:residue_N10} (a) illustrates the impurity residue upon varying the impurity-bath couplings $g_{AC}$ and $g_{BC}$. In all cases we find that $Z$ exhibits a broad peak around $Z=1$ with respect to $g_{AC}$ and decreases for finite positive and negative $g_{AC}$, implying the dressing of the impurity and thus its emergent polaronic character \cite{beyofro6}.
The reduction of $Z$ towards zero for strong repulsions $g_{AC}$ and $g_{BC}$ is caused by the phase-separation of the impurity with its hosts \footnote{For large $g_{AC}$ and $g_{BC}<0$, host A (core) phase-separates with the impurity (shell), while host B forms a less pronounced shell structure, leading to a minor dressing of the impurity.}, thus rendering no dressing possible, see \cite{supplemental_material}. In contrast, for strong attractions $g_{AC}\ll0$ the impurity either lies within both hosts for $g_{BC}<0$ or solely resides within host A while host B forms a shell structure for $g_{BC}>0$ \cite{supplemental_material}. In the latter case the density of host B localizes towards the trap edges, thereby encapsulating the other two species which reside around the trap center forming the core. As a result of the strong binding of the impurity with at least one of the baths and the potential phase separation with the species B, the impurity residue decreases towards zero. The width of the residue distribution for $g_{BC}=0$ is larger than for the binary system due to the smaller interspecies coupling strength $g_{AB}=0.1$, as compared to the intraspecies interaction $g=0.2$. For repulsive $g_{BC}$ we observe a decrease and shift of the impurity residue peak with increasing $g_{BC}$ as compared to the case of $g_{BC}=0$ and the binary mixture. These phenomena are again attributed to the presence of the impurity-hosts phase-separation taking place for a larger range of values of $g_{AC}$ with increasing $g_{BC}$, see Ref. \cite{supplemental_material}. Furthermore, the width of the residue distribution decreases for increasingly repulsive values of $g_{BC}$. Note the sudden decrease and subsequent increase of $Z$ for $g_{BC}=2.0$ and attractive $g_{AC}\simeq-1.8$ which is an imprint of a weak phase-separation of the impurity and the host A with species B, see Ref. \cite{supplemental_material}.
However, for attractive $g_{BC}$ a broadening of the residue distribution occurs towards repulsive values of the couplings $g_{AC}$, while the corresponding value of the maximum. The fact that $Z<1$ for $g_{AC}=0$ is attributed to the finite coupling $g_{BC}$, leading already to polaron formation by the host B, while for $g_{AC}\neq0$ the polaron experiences an additional dressing. 
Importantly, a powerful asset of the binary host is that depending on the combination of attractive and repulsive impurity-bath couplings it is possible to flexibly control and maintain the polaron for larger values of the coupling strength $g_{AC}$ to the medium. 
This effect can also be retrieved for heavier impurities, e.g. with mass ratios $m/m_C=87/133$ and $m/m_C=87/174$, where the baths consist of $^{87}\rm{Rb}$ atoms and the impurities are either $^{133}\rm{Cs}$ or $^{174}\rm{Yb}$ atoms, respectively, see Ref. \cite{supplemental_material}. 

The competition of the impurity-medium coupling strengths also naturally impacts the polaronic energy $\langle\hat{H}_C\rangle=\langle\Psi|\hat{H}_C|\Psi\rangle-\langle\Psi_0|\hat{H}_C|\Psi_0\rangle$ [Fig. \ref{fig:residue_N10}(b)]. While for increasingly attractive $g_{AC}$ an increase of the energy occurs for arbitrary values of $g_{BC}$, for repulsive $g_{AC}$ the energy tends to saturate towards different values depending on $g_{BC}$, see also below. For $g_{BC}>0$ and $g_{AC}>0$ we generally encounter larger polaron energies suggesting an increasing effective mass \cite{beyofro6} as compared to the case $g_{BC}<0$. Consequently, it is possible to distinguish between repulsive and attractive impurity-bath coupling strengths $g_{BC}$ based on the corresponding polaron energy.
Next, we estimate the polaron binding energy
\begin{equation}
\Delta E = E(N_C,g_{AC},g_{BC})-E(N_C=0,g_{AC}=0,g_{BC}=0)
\end{equation}
being defined as the energy difference due to the injection of the impurity, where $E(N_C,g_{AC},g_{BC})$ is the total energy of the system with $N_C$ impurities interacting with an effective strength $g_{AC}$ and $g_{BC}$ with the respective species [Fig. \ref{fig:residue_N10} (c)]. As expected, $\Delta E$ decreases for increasingly attractive $g_{AC}$ and saturates for repulsive values, similarly to the behavior of $\langle\hat{H}_C\rangle$. The former can be associated with a strong binding of the impurity to its hosts, thereby reducing $\Delta E$, while the latter is a consequence of the resultant phase-separation process where the impurity forms a shell structure around the baths \cite{supplemental_material,zinner_pol}. Evidently, we find a clear hierarchy of $\Delta E$ depending on $g_{BC}$, namely decreasing $g_{BC}$ apparently leads to a reduction of $\Delta E$ for any fixed $g_{AC}$. Therefore, experimentally, e.g. utilizing a radiofrequency scheme \cite{rf1,rf2,rf3}, the corresponding polaronic resonances are well distinguishable from each other. We have verified that a similar behavior of $\Delta E$ takes place for a larger cumulative bath with $N_A=N_B=50$ particles \cite{supplemental_material}.

To offer an intuitive understanding into the impurity's state for varying $g_{AC}$ and $g_{BC}$ we construct an effective potential \cite{ortho} by considering the Bose-Bose mixture as a static potential superimposed to the harmonic confinement of the impurity. It reads
\begin{equation}
V_{eff}=\frac{1}{2}m_C \omega^{2} x^2+g_{AC}\rho^{(1)}_A(x)+g_{BC}\rho^{(1)}_B(x),
\end{equation}
where $\rho^{(1)}_\sigma(x)$ is the one-body density of $\sigma=A, B$ bath species calculated within the correlated MB approach and thereby includes all necessary correlations, 
Accordingly, the impurity may occupy the eigenstates $|\phi_i\rangle$ of $V_{eff}$.
\begin{figure}[t]
	\includegraphics[scale=0.17]{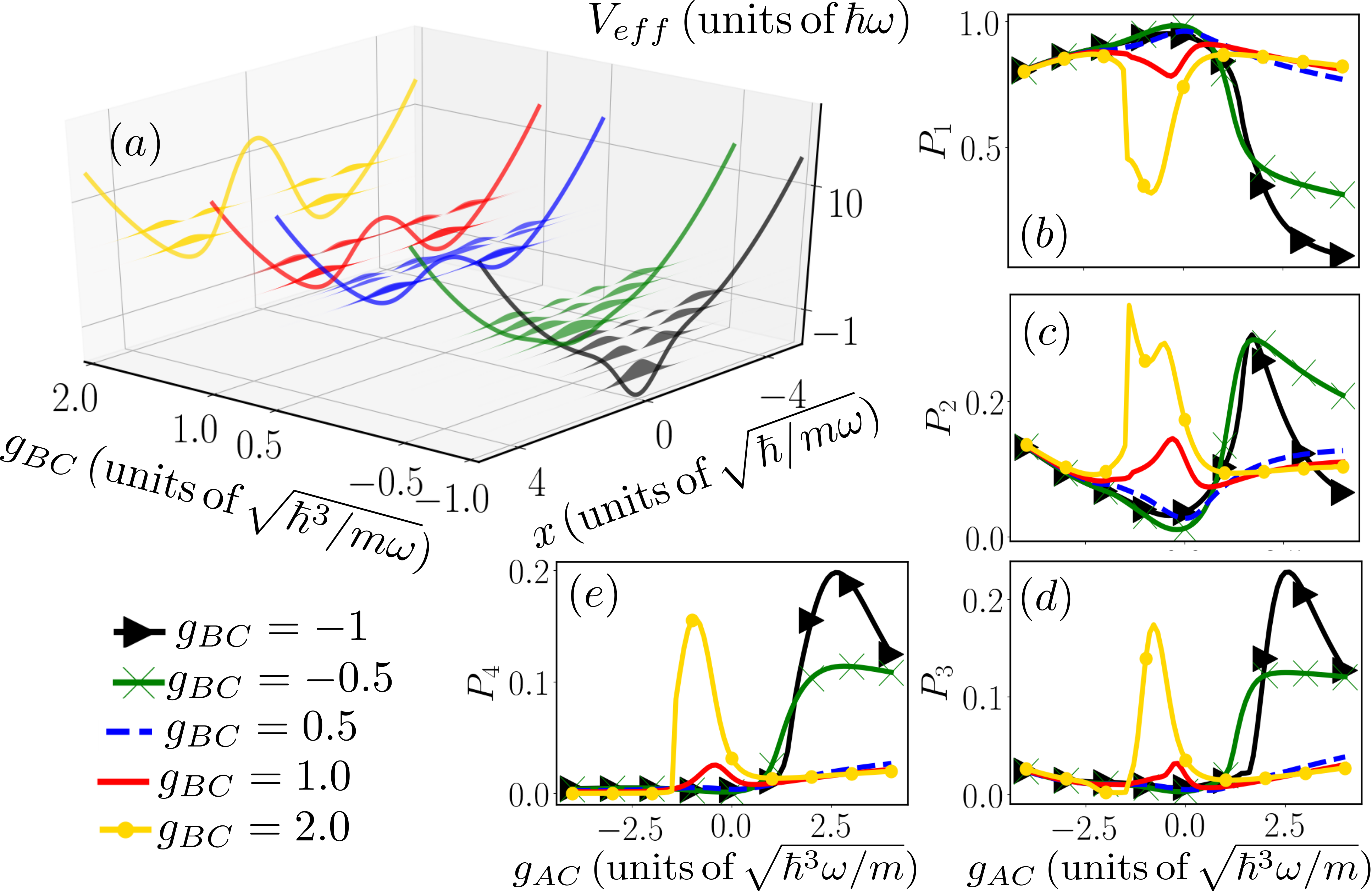}
	\caption{\label{fig:eff_pot}(a) Effective potential $V_{eff}$ and corresponding eigenvector distributions $|\phi_i(x)|^2$ for $g_{AC}=0.5$ and different $g_{BC}$. (b)-(e) Probability of finding the impurity in the single-particle eigenstate $|\phi_i\rangle$, $i=1,2,3,4$, of $V_{eff}$.}
\end{figure}
$V_{eff}$ neglects several phenomena such as the renormalization of the impurity's mass as well as the possible emergence of induced interactions.
Figure \ref{fig:eff_pot} (a) shows the deformations of the effective potential and its underlying eigenstates $|\phi_i\rangle$ under variations of $g_{BC}$ for $g_{AC}=0.5$. For $g_{BC}\ll0$, specifically $g_{BC}=-1$, the harmonic oscillator potential exhibits an additional dip which becomes more prominent with decreasing $g_{BC}$, whereas for $g_{BC}>0$ a double well structure forms. This has an impact on the related eigenstates such that quasi-degeneracies develop. The probability of finding the impurity in the $i$th eigenstate of $V_{eff}$ irrespectively of the states that are populated by the Bose-Bose mixture is given by
\begin{equation}
P_i=\sum_{kl}|\langle\vec{n}^A_k|\langle\vec{n}^B_l|\langle\phi_i|\Psi\rangle|^2,
\label{prob_imp}
\end{equation}
where $\{|\vec{n}^\sigma\rangle\}$ is an arbitrary complete Fock basis of the $\sigma=A, B$ baths. For all $g_{BC}$, except for $g_{BC}=2.0$, the ground state of the impurity is well described, i.e. $P_1>0.9$, by the corresponding ground state within the effective potential for weak attractive and weak repulsive $g_{AC}$ [Fig. \ref{fig:eff_pot} (b)]. Further decreasing $g_{AC}$ towards attractive couplings the occupation of $|\phi_1\rangle$ is reduced, whereas $|\phi_2\rangle$ starts to contribute [Fig. \ref{fig:eff_pot} (c)]. This behavior can still be recovered for $g_{AC}>0$, while for attractive impurity-bath couplings $g_{BC}<0$ and $g_{AC}>0$ we find a drastic decrease of $P_1$ accompanied by the substantial occupation of the excited states $|\phi_2\rangle$, $|\phi_3\rangle$ and $|\phi_4\rangle$ [Fig. \ref{fig:eff_pot} (c)-(e)]. Accordingly, the effective potential picture is no longer valid and does not provide a proper description of the impurity coupled to a cumulative bath. This is in line with the behavior of the impurity residue $Z$ which drops to zero in this interaction range [Fig. \ref{fig:residue_N10} (a)].
\begin{figure}[t]
	\includegraphics[scale=0.16]{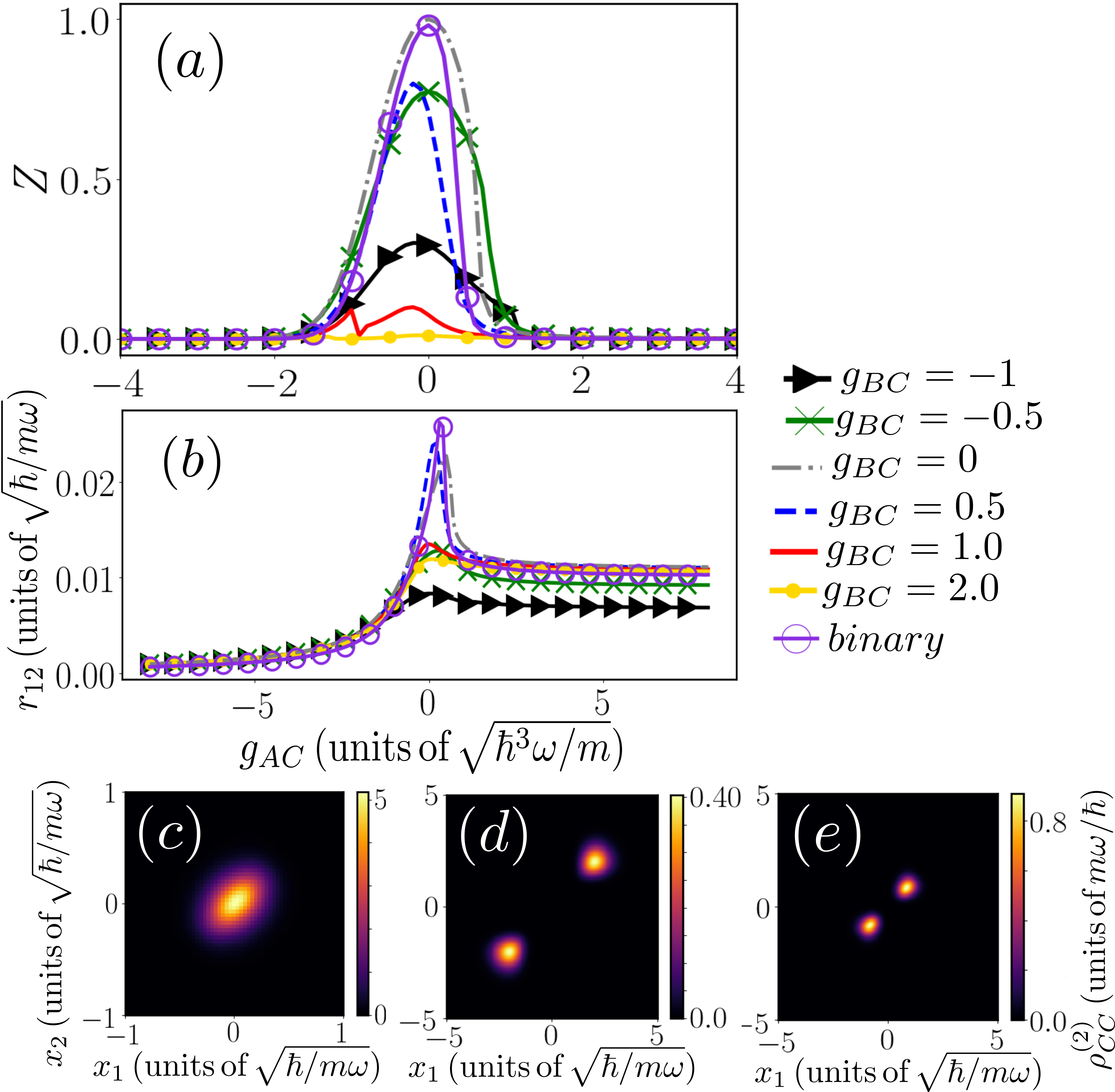}
	\caption{\label{fig:two_imp_residue_distance} (a) Two-impurity residue $Z$ and (b) distance $r_{12}$ for $N_C=2$ upon varying the impurity-bath couplings $g_{AC}$ and $g_{BC}$, while $g_{CC}=0$. The violet circles represent the binary mixture with $N=20$ bath atoms. Two-body density $\rho^{(2)}_{CC}(x_1,x_2)$ of the two impurities for combinations $(g_{AC},g_{BC})$ of (c) $(-2.1,-1)$, (d) $(8,2)$ and (e) $(8,-1)$.}
\end{figure}

Let us now discuss the behavior of $N_C=2$ bosonic impurities immersed in a Bose-Bose mixture. Monitoring the two-impurity residue [Eq. (\ref{eq:resdiue})] in order to extract the impact of the coupling on the impurities, we generally observe a similar behavior [Fig. \ref{fig:two_imp_residue_distance} (a)] as compared to the single impurity case. For $g_{BC}<0$ the peaks of $Z$ are broadened while being reduced in height, whereas for $g_{BC}>0$ a smaller width of the residue distribution is encountered compared to $N_C=1$. The effect of the additional impurity can be evinced in the strong suppression of the peak height of $Z$ for all finite impurity-bath couplings $g_{BC}$ (e.g. a reduction by $\sim 50\%$ for $g_{BC}=-1$), signalling stronger coherence losses when compared to the $N_C=1$ scenario \cite{rf2}. Consequently, this leads to a decreasing two-impurity residue $Z$.
In the case of $N_C=2$ the question regarding their effective interactions mediated by the hosts and being naturally related to their relative distance arises. This cannot be inferred from $Z$. For this reason we analyze the experimentally tractable \cite{selim_distance} impurity distance \cite{simosfermi,keiler3,bipolaron3}
\begin{equation}
r_{12}=\int\int dx_1dx_2|x_1-x_2|\rho_{CC}^{(2)}(x_1,x_2),
\label{eq:imp_distance}
\end{equation}
with  $\rho_{CC}^{(2)}(x_1,x_2)=\langle\Psi|\hat{\Psi}_{C}^{\dagger}(x_1)\hat{\Psi}_{C}^{\dagger}(x_2)\hat{\Psi}_{C}(x_1)\hat{\Psi}_{C}(x_2)|\Psi \rangle$ being the two-body density of two impurities providing the probability of finding simultaneously one impurity at position $x_1$ and the other one at $x_2$.
Around $g_{AC}=0$ we find a peak of the impurities' distance, which is most pronounced for $g_{BC}=0$, $g_{BC}=0.5$  and the binary system [Fig. \ref{fig:two_imp_residue_distance} (b)]. In all cases the manifestation of attractive induced interactions mediated by the hosts is evident by the decreasing behavior of $r_{12}$ for finite $g_{AC}$. Accordingly, the strength of the induced interactions in the region of the existence of the polaron becomes stronger when considering two hosts. More precisely $r_{12}$ features a decreasing trend towards zero for $g_{AC}<0$, while for $g_{AC}>0$ it saturates to a finite value. These finite values of $r_{12}$ barely differ from each other for $g_{BC}\geq0$ and in the case of the binary system. In sharp contrast, for $g_{BC}<0$ a saturation towards smaller distances is observed, indicating that the impurities lie closer with respect to one another. 

In order to clarify whether indeed induced interactions are established, we further investigate the impurities' two-body spatial distribution.
For triple mixture settings the structure and strength of the induced interactions are completely unexplored. Let us first discuss the interaction regime in which $r_{12}$ is independent of $g_{AC}<0$  [Fig. \ref{fig:two_imp_residue_distance} (b)]. As a characteristic example we present $\rho_{CC}^{(2)}$ for $g_{AC}=-2.1$ and $g_{BC}=-1$  [Fig. \ref{fig:two_imp_residue_distance} (c)]. Here, the two impurities lie together at the trap center and the probability to be located at different positions is reduced, yielding an elongated pattern along $x_1=x_2$. Hence, the impurities experience an induced interaction due to the cumulative bath. Importantly, this shrinking along the anti-diagonal of $\rho_{CC}^{(2)}$ is indicative  of a bound state having formed between the impurities known as a bipolaron state \cite{bipolaron1,bipolaron2,bipolaron3}. We remark that a bound bipolaron can only be formed for sufficiently attractive $g_{AC}$ in the case of $g_{BC}\geq0$ since the impurities' binding energy is not negative otherwise (not shown). However, for $g_{BC}<0$ the range of existence for a bound bipolaron extends towards weak repulsive couplings $g_{AC}$, e.g. around $g_{AC}\approx1$ for $g_{BC}=-1$. Turning now to the case of $g_{AC}>0$ for $g_{BC}=2$ [Fig. \ref{fig:two_imp_residue_distance} (d)] it is possible to infer that the impurities form a shell structure, indicating a phase-separation with their hosts. Moreover, they tend to occupy the same position, residing in a particular side of the appearing shell \cite{zinner_ind}. A slight elongation as for $g_{BC}=-1$ and $g_{AC}=-2.1$ [Fig. \ref{fig:two_imp_residue_distance} (c)] is also visible. Apart from forming a smaller shell structure for $g_{BC}=-1$ and $g_{AC}=8$ [Fig. \ref{fig:two_imp_residue_distance} (e)] the off-diagonal contribution is suppressed as compared to $g_{BC}=2$ and $g_{AC}=8$ [Fig. \ref{fig:two_imp_residue_distance} (d)], indicating the enhancement of the impurities induced interactions. This explains the saturation of $r_{12}$ towards a smaller value as compared to $g_{BC}=2$ [Fig. \ref{fig:two_imp_residue_distance} (b)]. In this sense, it is possible to steer the strength of the induced interactions by varying $g_{BC}$ as well as the width of the shell structure formed by the impurities.  

\paragraph{Conclusions.-}
Our results pave the way for controlling the quasiparticle character and induced interactions as well as to expose the role of correlations in triple mixture settings. The latter lays the foundations for studying related quantum phase transitions and pattern formation. Considering two impurity species trapped in a lattice and immersed in a medium, repulsively bound bipolarons of two species might be realized \cite{bipolaron4,imp_lattice_bath}. %The generalization of our findings to higher-dimensional settings, where topological effects play an important role, is certainly a valuable perspective.

Another intriguing step would be to consider the sudden injection of the impurity species \cite{ortho} into the Bose-Bose mixture for the simulation of experimental spectroscopic techniques \cite{beyofro2,rf1,rf2} in order to unravel the polaron dynamics. %In an experiment this quench protocol can be implemented by using a radiofrequency $\pi/2$ pulse on a spinor impurity species which is initially prepared in a non-interacting state with the cumulative bath. This transfers the impurities to a superposition of its interacting and non-interacting spin states. 
The corresponding spectral response provides characteristic information about the impurity residue and emergent dressed excited states, thus also offering a direct realization of our findings \cite{catani,Cetina_intef}.

\begin{acknowledgments}
The authors thank C. Weitenberg for a detailed feedback on the manuscript.
K. K. gratefully acknowledges a scholarship of the Studienstiftung des deutschen Volkes. S. I. M gratefully acknowledges financial support in the framework of the Lenz-Ising Award of the University of Hamburg.
\end{acknowledgments}

\clearpage

%%%%%%%%%% Merge with supplemental materials %%%%%%%%%%
\begin{widetext}
\begin{center}
	\textbf{\large Supplemental Material: Polarons and their induced interactions in highly imbalanced triple mixtures}
\end{center}
\end{widetext}
%%%%%%%%%% Merge with supplemental materials %%%%%%%%%%
%%%%%%%%%% Prefix a "S" to all equations, figures, tables and reset the counter %%%%%%%%%%
\setcounter{equation}{0}
\setcounter{figure}{0}
\setcounter{table}{0}
\setcounter{page}{1}
\makeatletter
\renewcommand{\theequation}{S\arabic{equation}}
\renewcommand{\thefigure}{S\arabic{figure}}
\renewcommand{\bibnumfmt}[1]{[S#1]}
\renewcommand{\citenumfont}[1]{S#1}
%%%%%%%%%% Prefix a "S" to all equations, figures, tables and reset the counter %%%%%%%%%%

\section{Variational many-body approach: ML-MCTDHX}
Our approach to determine the ground state properties of the triple mixture relies on the \textit{ab-initio} Multi-Layer Multi-Configuration Time-Dependent Hartree method for bosonic (fermionic) mixtures (ML-MCTDHX) \cite{mlb1,mlb2,mlx}, which accounts for all the relevant interparticle correlations \cite{Mistakidis_phase_sep,ff2018,ff2019,bf2018,lode1,lode2}.
As a first step, the total many-body (MB) wave function $|\Psi(t)\rangle$ is expanded in $M_\sigma$ species functions $|\Psi^{\sigma}(t)\rangle$ of species $\sigma$  
\begin{equation}
|\Psi(t)\rangle = \sum_{ijk=1}^{M_A,M_B,M_C} A_{ijk} |\Psi_{i}^A(t)\rangle\otimes |\Psi_{j}^B(t)\rangle \otimes |\Psi_{k}^C(t)\rangle,
\label{eq:psi}
\end{equation}
where the coefficients $A_{ijk}$ account for interspecies correlations \cite{horo}.
Furthermore, in order to capture the correlations within each component the species wave functions $|\Psi^{\sigma}(t)\rangle$ describing an ensemble of $N_\sigma$ bosons are expanded in a set of permanents
\begin{equation}
|\Psi_{i}^\sigma(t)\rangle = \sum_{\vec{n}^\sigma|N_\sigma} C_{\sigma\vec{n}}(t).
|\vec{n}^\sigma;t\rangle,
\label{eq:ml_ns}
\end{equation}
Here, the vector $\vec{n}^\sigma=(n^{\sigma}_1,n^{\sigma}_2,...)$ denotes the occupations of the time-dependent single-particle functions of the $\sigma$ species. The notation $\vec{n}^\sigma|N_\sigma$ indicates that for each $|\vec{n}^\sigma;t\rangle$ we require the condition $\sum_{i}n^{\sigma}_i=N_\sigma$.
The time propagation of the MB wave function is achieved by employing the Dirac-Frenkel variational principle $ \langle\delta\Psi| (\textrm{i}\partial_t - \mathcal{H} )|\Psi\rangle $ \cite{var1,var2,var3} with the variation $\delta\Psi$.
ML-MCTDHX provides access to the complete MB wave function of the triple mixture which consequently allows us to derive all relevant characteristics of the underlying system. As such we are able, among others, to characterize the system by projecting onto number states with respect to an appropriate single-particle basis \cite{ns_analysis1,ns_analysis2}. Besides investigating the quantum dynamics it allows us to determine the ground (or excited) states by using either imaginary time propagation or improved relaxation \cite{meyer_improved}, thereby being able to uncover also possible degeneracies of the involved MB states. We remark that in commonly used approaches for solving the time-dependent Schr{\"o}dinger equation, one typically constructs the wave function as a superposition of time-independent Fock states with time-dependent coefficients. Instead, it is important to note that the ML-MCTDHX approach considers a co-moving time-dependent basis on different layers, meaning that in addition to time-dependent coefficients the single particle functions spanning the number states are also time-dependent. This leads to a significantly smaller number of basis states and configurations that are needed to obtain an accurate description of the system under consideration and thus renders the treatment of mesoscopic systems feasible \cite{pruning}. 

The degree of truncation of the underlying Hilbert space is given by the orbital configuration $C=(M_A,M_B,M_C,d_A,d_B,d_C)$. Here, $M_\sigma$ refers to the number of species functions in Eq. \ref{eq:psi}, while $d_\sigma$ with $\sigma \in\{A,B,C\}$ denote the number of single-particle functions spanning the time-dependent number states $ |\vec{n}^\sigma; t\rangle$ (cf. equation \ref{eq:ml_ns}). The orbital configuration $C=(6,6,6,4,4,6)$ has been employed for all MB calculations presented in the main text, yielding a converged behavior of our observables.

\section{Correlations of the triple mixture and emergent phase-separation in the spatial distributions}

Further insight into the underlying processes related to the polaron properties can be gained by analyzing the spatial distribution of the three species in terms of the one-body density of the ground state $|\Psi\rangle$ of the species $\sigma=A,B,C$, which is defined as
\begin{equation}
\rho^{(1)}_\sigma(x)=\langle\Psi| \hat{\Psi}_{\sigma}^{\dagger}(x)\hat{\Psi}_{\sigma}(x)|\Psi \rangle.
\label{eq:obd}
\end{equation}
The spectral decomposition of the one-body density of species $\sigma$ reads
\begin{equation}
\rho_\sigma^{(1)}(x) = \sum_j n_{\sigma j} \Phi^{*}_{\sigma j}(x)\Phi_{\sigma j}(x),
\label{eq:natural_populations}
\end{equation}
where $n_{\sigma j}(t)$ in decreasing order, obeying $ \sum_j n_{\sigma j}=1$, are the so-called natural populations and $\Phi_{\sigma j}(x,t)$ the corresponding natural orbitals. In this sense, the $\sigma$-species natural orbitals are the eigenstates, while the natural populations are the corresponding eigenvalues \cite{meyer_improved}, which are determined by diagonalizing the $\sigma$-species one-body reduced density matrix. The natural populations serve as a measure for the correlations in a subsystem. Accordingly, in order to quantify the degree of correlations or fragmentation we resort to the $\sigma$-species entropy \cite{lode1,lode2,Keiler_tunel} defined as
\begin{equation}
S_{\sigma}(t)=-\sum_{j} n_{\sigma j}(t) \ln(n_{\sigma j}(t)).
\label{eq:fragmentation}
\end{equation}
Here, the case of $S_{\sigma}=0$ indicates that the subsystem $\sigma$ is not depleted, implying that all particles occupy the same single particle state, i.e. $n_{\sigma 1}=1$.
\begin{figure}[t]
	\includegraphics[scale=0.2]{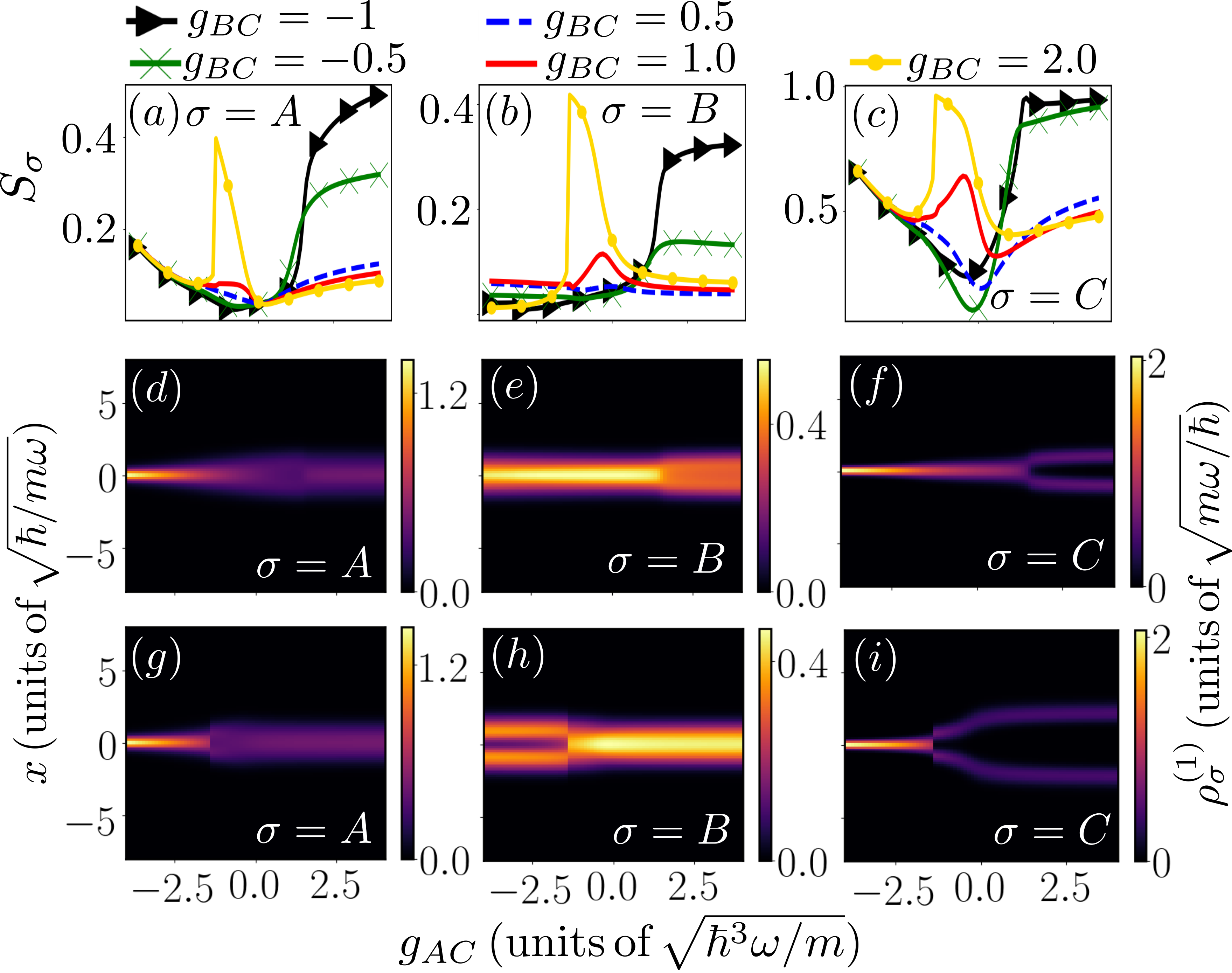}
	\caption{\label{fig:densities_correlations} (a) - (c) Entropy $S_\sigma$ for varying $g_{AC}$ and $g_{BC}$ quantifying the degree of intraspecies correlations. (d)-(f) One body density $\rho_\sigma^{(1)}(x)$ for $g_{BC}=-1$ (middle row) and $g_{BC}=2$ (lower row) for different impurity-bath couplings $g_{AC}$, showcasing that e.g. for $g_{AC}=-3$, $g_{BC}=-1$ the impurity lies in their hosts, while for $g_{AC}=3$, $g_{BC}=-1$ a phase-separation occurs. Here $g_{AA}=g_{BB}=0.2$, $g_{AB}=0.1$, $m_A=m_B=m_C$, $N_A=N_B=10$ and $N_C=1$ .}
\end{figure}
Fig. \ref{fig:densities_correlations} (a)-(c) shows the fragmentation $S_\sigma$ for the respective species in the case of $N_C=1$ and $N_A=N_B=10$. For $g_{BC}>0$ we observe a minor increase (decrease) of $S_A$ ($S_B$) towards repulsive and attractive values of $g_{AC}$, whereas the impact on $S_C$ is more pronounced. Attractive impurity-host couplings $g_{BC}$ lead to a drastic increase of all $S_\sigma$ for repulsive $g_{AC}$. This behavior of the fragmentation can also be observed in the context of the occupation of the states $|\Phi_i\rangle$ in the effective potential $V_{eff}$  [see Fig. \ref{fig:eff_pot}].
Due to the emergent strong correlations a correlated approach, such as ML-MCTDHX, is needed in order to properly describe the system.

Moreover, the existence of correlations is imprinted in the spatial distribution of the mixture and the impurity. Recall that for a single impurity interspecies correlations (entanglement) between the impurity and the cumulative bath (Bose-Bose mixture) are accounted for by $S_C$. For strongly repulsive $g_{AC}$ the impurity forms a shell structure [Fig. \ref{fig:densities_correlations} (f),(i)] and a phase-separation can be observed. The separation of the shell is smaller for attractive $g_{BC}$ as compared to repulsive ones due to the attraction to the B species. This attraction also leads to the formation of a less pronounced shell structure in the B species and thereby enhances the overlap with the impurity. We can interpret this as a weak phase-separation between the A species and the B species as well as the impurity species with both hosts [Fig. \ref{fig:densities_correlations} (d)-(f)]. As a result the fragmentation is significantly increased. However, for $g_{BC}>0$ and large repulsive $g_{AC}$ the phase-separation takes place between the impurity and the cumulative bath, while the two hosts remain miscible [Fig. \ref{fig:densities_correlations} (g)-(i)], a process that leads to a smaller fragmentation in all cases. 
The peak in the relevant entropic measures for $g_{BC}=2$ and weakly attractive $g_{AC}$ is due to the formation of a shell structure in the A species and the impurity species, similar to the case of $g_{BC}<0$ and strongly repulsive $g_{AC}$.

\section{Polaron properties in the mean-field framework}
\begin{figure}[t]
	\includegraphics[scale=0.2]{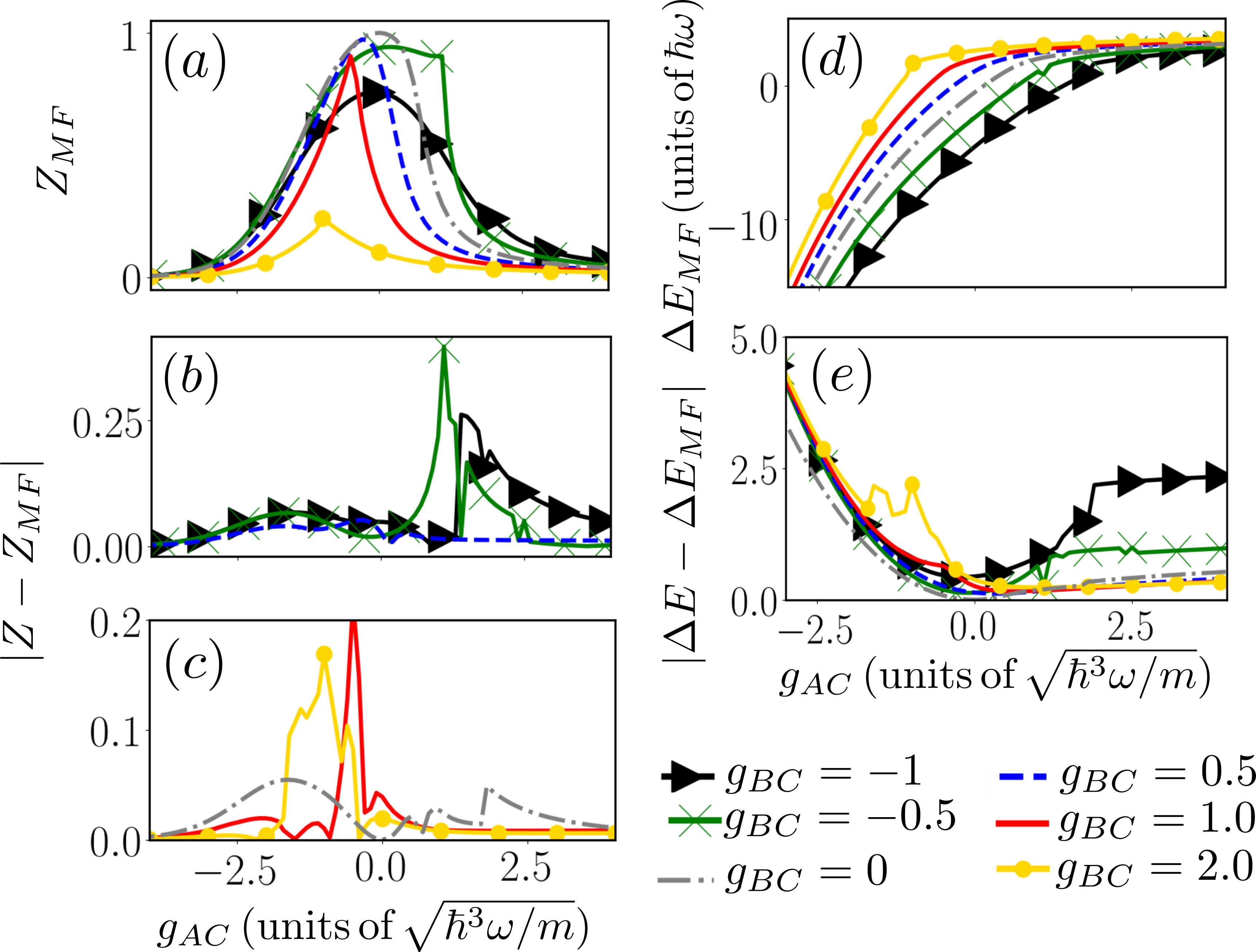}
	\caption{\label{fig:residue_MF} Polaron (a) residue $Z$, (d) binding energy $\Delta E$ for different impurity-bath couplings $g_{AC}$ and $g_{BC}$ employing a mean-field ansatz. Absolute difference for (b), (c) the residue $|Z-Z_{MF}|$ and (e) the deviation of the binding energy $|\Delta E- \Delta E_{MF}|$ between employing the MF ansatz and the MB treatment for varying $g_{AC}$ and $g_{BC}$. The same system parameters as in Fig. \ref{fig:densities_correlations} are used.}
\end{figure}
As it becomes evident from Eqs. \ref{eq:psi}, \ref{eq:ml_ns}, ML-MCTDHX is able to operate within different approximation orders. For instance, it reduces to the set of coupled mean-field (MF) Gross-Pitaevskii equations of motion when $C=(1,1,1,1,1,1)$. Moreover, in case that $A_{111}=1$ the species $A$, $B$ and $C$ are not entangled \cite{horo} but intraspecies correlations can be taken into account. As a result the system is described within a species mean-field approximation (SMF) corresponding to a single product state ansatz, characterized by $M_A=M_B=M_C=1$ \cite{theel1}.
In the following we aim to reveal the necessity of a fully correlated approach, i.e. accounting for all the emergent intra- and interspecies correlations, for determining the ground state of the polaron. In this sense, as we shall demonstrate a standard MF ansatz is not sufficient for describing the polaronic properties e.g. discussed in Fig \ref{fig:residue_N10}. To validate this assumption, we subsequently determine the ground state employing a MF ansatz, $C=(1,1,1,1,1,1)$, and calculate the polaron residue as well as its binding energy [Fig. \ref{fig:residue_MF}]. Qualitatively we find a similar behavior of the residue distribution $Z_{MF}$ [Fig. \ref{fig:residue_MF} (a)], using a MF ansatz, for the different impurity-medium couplings $g_{AC}$ and $g_{BC}$ as compared to the residue where the underlying MB wavefunction, taking correlations into account, has been considered [see Fig. \ref{fig:residue_N10} (a)]. Hence, the broadening of the residue distribution for $g_{BC}<0$ which effectively leads to a stabilization of the polaronic character for larger interactions can also be predicted using a MF ansatz. However, upon investigating the absolute difference $|Z-Z_{MF}|$ we quantitatively identify strong deviations between the two distributions for various $g_{AC}$ and $g_{BC}$ [Fig. \ref{fig:residue_MF} (b),(c)]. Consequently, the MF ansatz over- or underestimates the residue distribution at specific interaction intervals. For strong repulsive or attractive impurity-host couplings $g_{AC}$ a comparison is not adequate since in both cases, i.e. MF and MB, there is a phase-separation of the impurity with their hosts.
Turning now to the polaron binding energy, employing a MF ansatz [Fig. \ref{fig:residue_MF} (d)] we find a clear hierarchy of the polaronic resonances as already discussed in Fig. \ref{fig:residue_N10} (c). Considering the absolute difference $|\Delta E- \Delta E_{MF}|$ with respect to the MB treatment we capture also here a quantitative deviation between the two approaches [Fig. \ref{fig:residue_MF} (e)]. In particular, for $g_{AC}<0$ the deviation $|\Delta E- \Delta E_{MF}|$ increases with decreasing $g_{AC}$, while for repulsive $g_{AC}$ it saturates towards a finite value with increasing $g_{AC}$. Only for weak impurity-bath couplings the MF approach is able to reproduce binding energies which are close to the ones using a full MB treatment.
In case of a SMF approximation we find deviations $|Z-Z_{SMF}|$ and $|\Delta E- \Delta E_{SMF}|$ which are very similar to the ones observed for a MF ansatz (not shown here for brevity). This evinces that indeed interspecies correlations play a crucial role in the polaron formation.

\section{Heavy impurities and larger hosts}
 \begin{figure}[t]
	\includegraphics[scale=0.2]{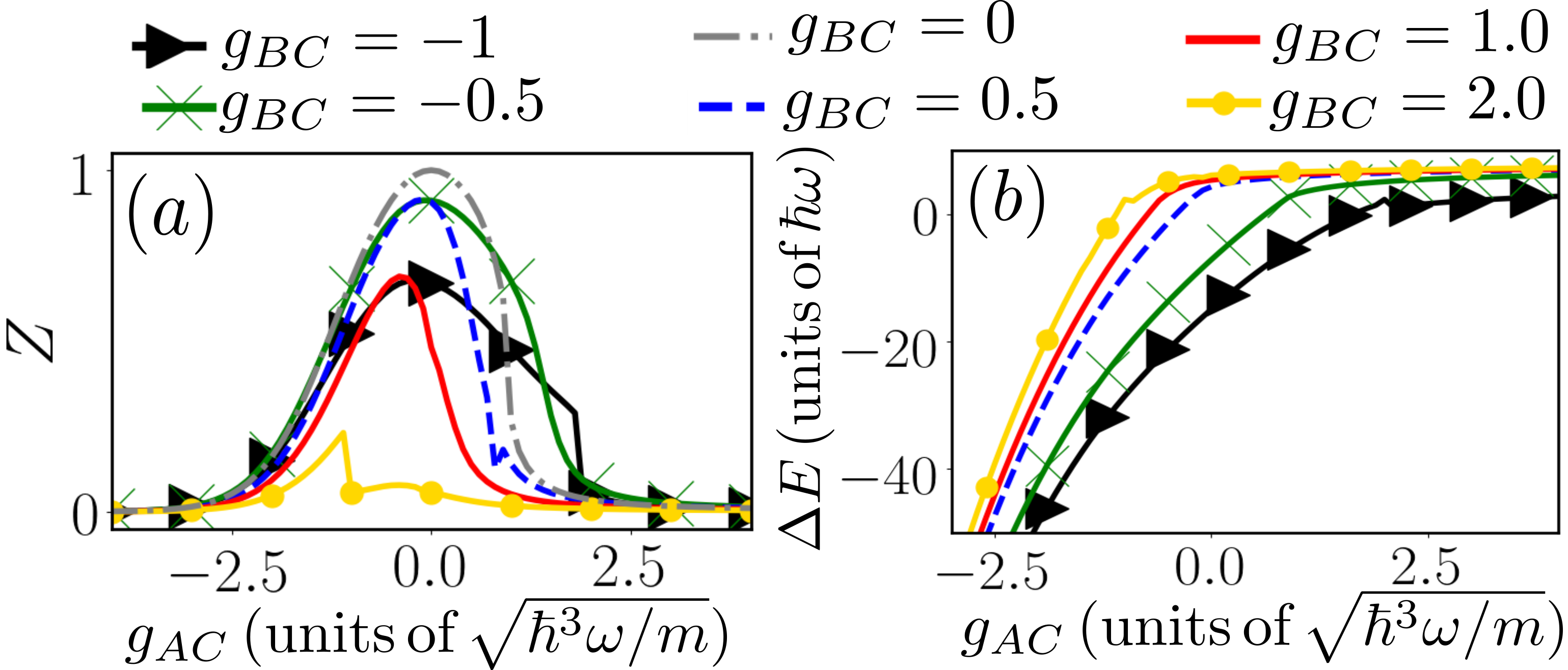}
	\caption{\label{fig:residue_Yt} Polaron (a) residue $Z$ for $m_A/m_C=87/174$ and $N_A=N_B=10$, (b) binding energy $\Delta E$ for $m_A/m_C=1$ and $N_A=N_B=50$ upon varying the impurity-bath couplings $g_{AC}$ and $g_{BC}$. The other system parameters are the same as in Fig. \ref{fig:densities_correlations}.}
\end{figure}
Our findings regarding e.g. the polaron residue are not limited to the case of a mass-balanced triple mixture, i.e. $m_A=m_B=m_C$ [Fig. \ref{fig:residue_N10} (a)], but can also be generalized for heavier impurities, namely $^{133}\rm{Cs}$ or $^{174}\rm{Yb}$ such that $m/m_C=87/133$ and $m/m_C=87/174$, respectively. As an example we will discuss the case of $^{174}\rm{Yb}$ and determine the residue according to Eq. \ref{eq:resdiue}. Fig. \ref{fig:residue_Yt} (a) presents $Z$ upon variation of the impurity-medium coupling strengths $g_{AC}$ and $g_{BC}$. Qualitatively, the residue distribution is similar to the case of equal masses, while exhibiting slightly larger (smaller) widths for $g_{BC}<0$ ($g_{BC}>0$). We solely find minor quantitative deviations between a $^{174}\rm{Yb}$ and a $^{87}\rm{Rb}$ impurity. Importantly, this implies that also for heavier impurities the quasiparticle character can be maintained for larger couplings $g_{AC}$ when $g_{BC}<0$, while for $g_{BC}>0$ the width of the residue distribution with respect to $g_{AC}$ is smaller as compared to $g_{BC}=0$. 

State-of-the-art ultracold atom experiments are often of mesoscopic character and consist of $\simeq100$ particles.
For this reason, we compute the polaron binding energy for a larger cumulative bath with $N_A=N_B=50$ bosons [Fig. \ref{fig:residue_Yt} (b)]. Similar to the case of $N_A=N_B=10$ [Fig. \ref{fig:residue_N10} (c)], $\Delta E$ decreases towards attractive $g_{AC}$ and saturates for repulsive ones. The latter can again be attributed to the phase-separation of the impurity with respect to its hosts, where the impurity forms a shell-structure. The clear hierarchy of $\Delta E$ in terms of $g_{AC},g_{BC}$ can be retrieved for larger hosts, such that the polaronic resonances are well distinguishable from each other e.g. using radiofrequency schemes. Note that the polaron binding energies are strongly reduced, suggesting an increased effective mass, for a large range of impurity-medium couplings $g_{AC}$ as compared to $N_A=N_B=10$.

\section{Polarons in homogeneous settings}
\begin{figure}[t]
	\includegraphics[scale=0.2]{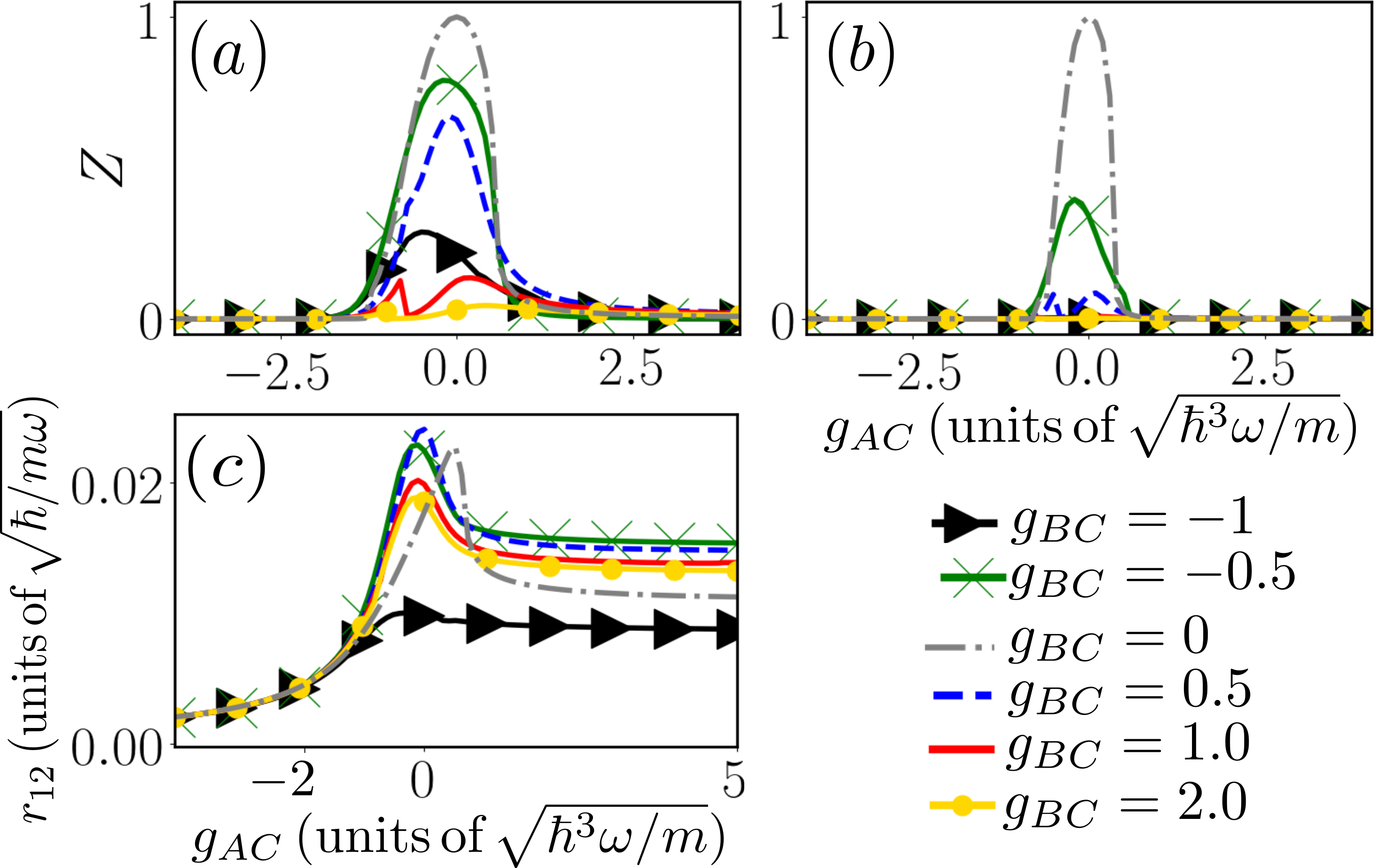}
	\caption{\label{fig:residue_box} Polaron residue $Z$ for (a) $N_C=1$, (b) $N_C=2$ and (c) impurity distance $r_{12}$ for $N_C=2$ upon varying the impurity-bath couplings $g_{AC}$ and $g_{BC}$. All species solely experience a box potential. The other system parameters are the same as in Fig. \ref{fig:densities_correlations}.}
\end{figure}
In our work we demonstrate the Bose polaron properties in a cumulative bath by explicitly accounting for trap effects in terms of a harmonic confinement. These are almost inevitable in contemporary experiments of cold atomic settings. To underline the importance of including an external trapping potential we examine its impact on the polaronic properties when considering solely a box potential. Specifically, we omit the terms $\frac{1}{2}m_\sigma \omega^{2}_\sigma x^2$, $\sigma \in \{A,B,C\}$, in the Hamiltonian [Eq. \ref{eq:hamiltonian}], while keeping the hard wall boundary conditions. The other system parameters remain the same to those utilized in the main text.

As a first step, we determine the polaron residue for a single impurity immersed in a Bose-Bose mixture upon varying the involved impurity-medium coupling strengths [Fig. \ref{fig:residue_box} (a)]. It can be readily seen that the distribution of $Z$ with respect to $g_{AC}$ is significantly reduced as compared to the trap scenario depicted in Fig. \ref{fig:residue_N10} (a). Indeed, for all $g_{BC}$ the peak height as well as the width of the residue distribution are strongly reduced. E.g. while in the trapped case [Fig. \ref{fig:residue_N10} (a)] for $g_{BC}=1$ the peak of the $Z$-distribution is rather pronounced, neglecting the harmonic confinement leads to a drastic decrease such that the peak is barely visible. Moreover, the increased width of the residue distribution for $g_{BC}<0$ as compared to $g_{BC}=0$ cannot be recovered. Hence, the polaron state cannot be maintained for larger impurity-medium couplings when considering solely a box potential.
The effect on $Z$ is even more dramatic for $N_C=2$ impurities [Fig. \ref{fig:two_imp_residue_distance} (b)]. Here, only for $g_{BC}=0$ and $g_{BC}=-0.5$ a polaron state exists when considering weak impurity-bath couplings $g_{AC}$ [Fig. \ref{fig:residue_box} (b)].  Interestingly, the impurity distance $r_{12}$ is less affected by the presence of the external potential [Fig. \ref{fig:residue_box} (c)]. Here, we qualitatively find a similar behavior to the case of a harmonic confinement [Fig. \ref{fig:two_imp_residue_distance} (c)]. Namely, $r_{12}$ features a decreasing trend towards zero for $g_{AC}<0$, while for $g_{AC}>0$ it saturates to a finite value for all $g_{BC}$. For $g_{AC}<0$ again a shrinking along the anti-diagonal of $\rho_{CC}^{(2)}$ appears, being indicative of bipolaron formation, while for $g_{AC}>0$ the impurities form a shell structure, indicating a phase-separation with their hosts (not shown here for brevity). Nevertheless, we observe a quantitative change of the values of $r_{12}$ when neglecting the harmonic confinement, resulting e.g. in an increase of the impurity distance for large $g_{AC}$ in the case of $g_{BC}=-0.5$.

\end{document}